\documentclass[11pt,twoside]{article}

\usepackage{asp2004}
\usepackage{epsf}
\usepackage{psfig}
\usepackage{lscape}
\usepackage{graphicx}

\begin{document}
\title{Unusual PAH Emission in Elliptical Galaxies}   
\author{Jesse D. Bregman\altaffilmark{2}, Joel N. Bregman\altaffilmark{1}, and Pasquale Temi\altaffilmark{2,3,4} }

\altaffiltext{1}{Department of Astronomy, University of Michigan, 
Ann Arbor, MI 48109}
\altaffiltext{2}{Astrophysics Branch, NASA Ames Research Center, 
MS 245-6, Moffett Field, CA 94035}
\altaffiltext{3}{SETI Institute, 515 N. Whisman Road, Mountain View, CA 94043.}
\altaffiltext{4}{Department of Physics and Astronomy, University of Western
Ontario, London, Ontario, N6A 3K7, Canada.}

\begin{abstract} 
In a sample of thirty normal elliptical galaxies observed with the Spitzer IRS, one galaxy, NGC4697, shows strong PAH emission, but with an apparently weak 7.7 \micron\ feature.  We find that the PAH emission is confined to the central regions of the galaxy and that once a quiescent elliptical galaxy spectrum is subtracted, the PAH feature ratios are normal.  We show that the PAH spectrum resembles the diffuse ISM of our galaxy rather than an HII region, and is not indicative of a starburst galaxy.  We suggest that the PAHs in NGC4697 are consistent with a recent but now past burst of star formation.
\end{abstract}


\section{Introduction} 
Before Spitzer, the few available mid-infrared spectra of elliptical galaxies only showed silicate emission from the shells around AGB stars \citep{athey}, so it was surprising when one of thirty normal elliptical galaxies from our Spitzer cycle 1 proposal (NGC4697) showed strong PAH emission in its IRS spectrum. \citet{kaneda} also found PAH emission in several galaxies chosen because they had bright X-ray and far infrared emission, and in two nearby galaxies from the SINGS program.  While one of these galaxies, NGC3265, showed PAH emission similar to a starburst galaxy, the others showed clear emission in the 11.3 \micron\ PAH emission feature with a weak or absent 7.7 \micron\ feature. \citet{kaneda} suggested that this could be due to emission from neutral PAHs rather than the ionized PAHs that produce the typical PAH spectrum.

In this paper we will show that the odd appearance of the PAH emission spectrum is due to a minimum near 8 \micron\ in the underlying continuum, and that the presence of PAHs suggest star formation in the recent past.

\section{Data Processing}

Data reduction started with the bcd data from the standard IRS data reduction pipeline at the Spitzer 
Science Center (SSC). 
Before performing the spectral extraction, the local background for the SL
modules was subtracted using observations when the target was located 
in an alternate slit. At longer wavelengths, since we recorded data using only 
the LL2 module, the local background was subtracted by differencing the 
two nod positions along the slit.
The spectra were then extracted from the sky-subtracted two-dimensional 
array images using the SMART software package \citep{higd04},
after the mean fluxes from each ramp cycle were
combined.
We performed a "fixed column extraction" because the emission from each
galaxy did not resemble the point source profile, showing substantial 
extended emission. 
In its current form, the SMART software is optimized to perform 
spectral extraction and flux calibration for point source targets.
In order to correct
for the use of the standard flux conversion tables, which are based
on point source extraction, we applied a
correction that accounts for the aperture loss due to the
narrowing of the extracting column as a function of wavelength used by 
SSC and that in turn feeds back into the FLUXCON tables.
Observations of the standard star HR 6348 were used to calibrate our
target spectra. A spectrum of HR 6348 was constructed combining a number
of observations recorded in 2004 under the program ID 1404.
Our spectra were calibrated by dividing the extracted spectrum of the
source by the spectrum of the standard star, extracted with the same
extraction parameters applied to our target sources, and multiplying
by its template \citep{cohen03}.

\section{Discussion}

The spectrum of the central region of NGC4697 is different than the outer region.  Figure 1 shows a 14 pixel wide extraction spectrum of NGC4697 (upper curve) and the spectrum of just the outer regions of the galaxy (lower curve).  The ratio of the PAH features in the full extraction are not typical of PAH spectra of objects within our galaxy or in starburst galaxies, with the 7.7 \micron\ feature appearing much weaker than normal.  In a typical quiescent elliptical galaxy, the spectrum consists of a 5-8 \micron\ stellar continuum and a stellar continuum with silicates superimposed longward of 8 \micron. The outer regions of NGC4697 resemble a typical elliptical galaxy with a weak 11.3 \micron\ PAH feature added, but the central region is dominated by PAH emission.  Clearly the PAH emission is concentrated towards the center of the galaxy.  However, as shown in Figure 2, once the underlying stellar+silicate emission is subtracted, the relative band ratios are more normal and closely resemble the spectrum of a cirrus knot in the diffuse ISM of our galaxy. The primary differentiator between these environments is not the PAH band ratios, but rather the shape of the underlying continuum.  Regions near bright UV sources (HII regions and reflection nebulae) show a continuum rising sharply longward of 15 \micron\, while the diffuse ISM spectrum (represented by a knot of Cirrus emission) does not show the long wavelength rise.

For the one galaxy observed by \citet{kaneda} to have PAH emission that is currently available in the Sptizer archive, NGC4589, subtracting an elliptical galaxy template reveals a 7.7 \micron\ PAH feature that is not apparent in the original data  (Figure 3).  The 11.3/7.7 \micron\ feature peak intensity ratio is about three in this galaxy, while in both the diffuse ISM of our galaxy and NGC4697, the ratio is close to one.  The weakness of the 7.7 \micron\ feature relative to the 11.3 \micron\ feature can be explained either as a predominance of neutral PAHs in NGC4589, as suggested by \citet{kaneda}, as a PAH mixture that is lacking small PAHs, or as the result of a cooler radiation field \citep{schutte}.  PAH ionization depends on the ratio of the UV field strength to the electron density.  At high values of this ratio, PAHs exist primarily as cations while at low values they are primarily anions.  PAHs will be neutral only in a narrow range of this ratio, and neutral PAHs may never constitute a large fraction of the PAH population \citep{bregman05}.  Neutral PAHs have a much higher 11.3/7.7 \micron\ feature ratio than either cations or anions.  The second way to increase the 11.3/7.7 \micron\ ratio is to reduce the number of small PAHs relative to the large ones.  Since the number of modes in a PAH molecule is N(N+1), where N is the number of carbon atoms in the molecule, and the energy from an absorbed photon is spread among all of the modes, the average excitation within a large PAH is lower than in a small PAH.  Thus, the higher energy states have a lower population in large molecules than in smaller ones.  \citet{schutte} showed that as the result of this effect, small PAHs contribute mostly to the short wavelength features and large molecules contribute mostly to the longer wavelength features.  If the small molecules are removed, the shorter wavelength features become weaker relative to the longer wavelength features.  The third possibility, a cooler radiation field, provides fewer high energy photons to excite the short wavelength modes.  Distinguishing which mechanism is responsible for these data is not possible with the current data set.

From the comparisons shown in Figure 2, it appears that the PAH emission in NGC4697 arises from PAHs in the diffuse ISM of the galaxy, not from a recent starburst.  However, the lifetime of PAH molecules is expected to be short in the hot ISM of elliptical galaxies ($\approx10^8$ years), so the PAHs must have been recently ejected into the ISM.  One possibility is that the spectral sequence from strong to weak PAH emission, NGC3265-$>$NGC4697-$>$NGC4589, is also a time sequence.  In NGC3265 a starburst is currently in progress, one has occurred in NGC4697 in the recent past, while in NGC4589 the remnants of the starburst are barely detectable.  Recently, \citet{sambhus} have noted that there are two different kinematic groups of planetary nebulae in NGC4697, perhaps indicating a recent merger event, and \citet{pinkney} conclude that NGC4697 is a candidate for a recent merger based on the presence of a central dusty disk.  This same event could have set off a burst of star formation, and we are now seeing the remnants of that event via PAH emission from the ISM of the galaxy.  

\acknowledgements 
We wish to thank the Spitzer Science Center and NASA for making available the support and observing time that made this project possible, and Dr. A. Tielens for valuable discussions.


\begin{figure}
\plotone{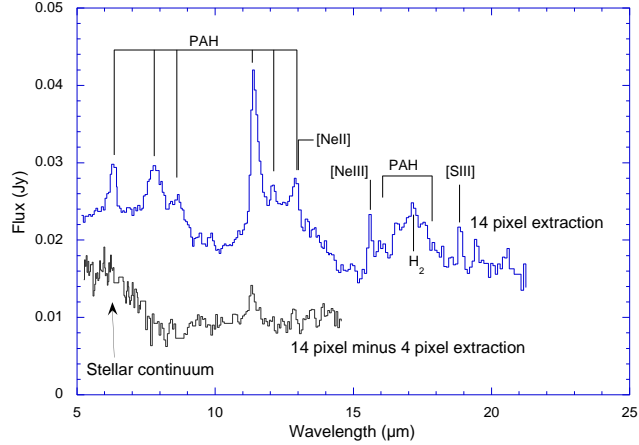}
\caption{The spectrum of NGC4697 shows strong PAH emission concentrated towards the center of the galaxy.  Shown here is the full galaxy spectrum (upper curve) and the spectrum with the central 4 pixel region excluded (bottom curve).  The bottom spectrum shows a stellar continuum from 5-8 \micron\, indicating that the observed full galaxy spectrum is a combination of a typical elliptical spectrum with PAH emission added.
}
\end{figure}

\begin{figure}
\plotone{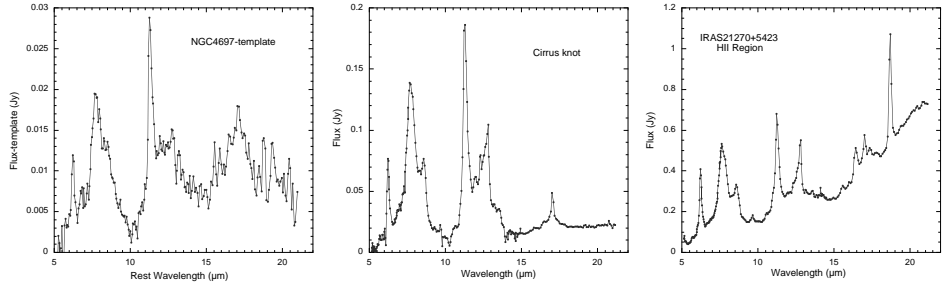}
\caption{These spectra compare the PAH emission in NGC4697 with different regions in our galaxy.  An elliptical galaxy template has been subtracted from the NGC4697 data to produce the spectrum of just the PAH emitting region (left panel).  The other panels show a bright cirrus knot in our galaxy (center panel) and the HII region IRAS21270+5423 (right panel).}
\end{figure}

\begin{figure}
\plotone{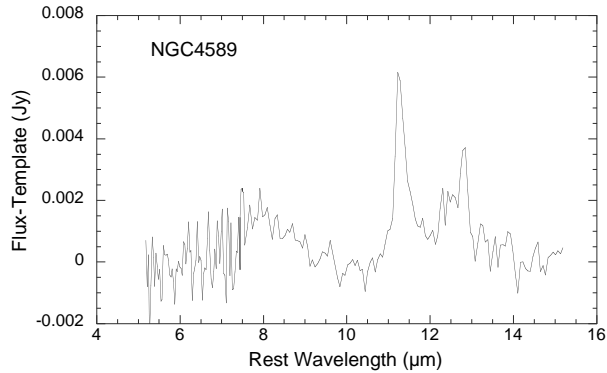}
\caption{The spectrum of NGC4589, one of the galaxies from \citet{kaneda}, has been subtracted from a quiescent elliptical galaxy template revealing the 7.7 \micron\ PAH emission feature.}
\end{figure}


\end{document}